\documentclass[runningheads]{llncs}
\usepackage[T1]{fontenc}
\usepackage{graphicx}
\usepackage{multirow}
\usepackage{arydshln}
\usepackage{xcolor}
\usepackage{xr}
\usepackage{hyperref}
\usepackage{subfiles} 
\usepackage{pdfpages}

\newcommand{\subpara}[1]{\vspace{0.1cm}\noindent\textbf{#1}}

\begin{document}
\title{Empirical Analysis of a Segmentation Foundation Model in Prostate Imaging}
\author{Heejong Kim\inst{1} \and
Victor Ion Butoi\inst{3}\and
Adrian V. Dalca\inst{3,4}\and
Daniel J.A. Margolis\inst{1}\and
Mert R. Sabuncu\inst{1,2}} 
\authorrunning{H. Kim et al.}
%
\institute{Department of Radiology, Weill Cornell Medicine, USA \and 
School of Electrical and Computer Engineering, Cornell University and Cornell Tech, USA
\and Martinos Center for Biomedical Imaging, Massachusetts General Hospital and Harvard Medical
School, USA \and Computer Science and Artificial Intelligence Laboratory, Massachusetts Institute of Technology, USA}

\titlerunning{Empirical Analysis of a Segmentation Foundation Model}

\maketitle              
\begin{abstract}
Most state-of-the-art techniques for medical image segmentation rely on deep-learning models. These models, however, are often trained on narrowly-defined tasks in a supervised fashion, which requires expensive labeled datasets. Recent advances in several machine learning domains, such as natural language generation have demonstrated the feasibility and utility of building foundation models that can be customized for various downstream tasks with little to no labeled data. This likely represents a paradigm shift for medical imaging, where we expect that foundation models may shape the future of the field. In this paper, we consider a recently developed foundation model for medical image segmentation, UniverSeg~\cite{butoi2023universeg}.
We conduct an empirical evaluation study in the context of prostate imaging and compare it against the conventional approach of training a task-specific segmentation model. Our results and discussion highlight several important factors that will likely be important in the development and adoption of foundation models for medical image segmentation.

\end{abstract}

\section{Introduction}
Foundation models (FMs) are general-purpose models trained on extensive amounts of data, typically in a self-supervised fashion~\cite{bommasani2021opportunities}. 
These pre-trained models can serve as the `foundation' from which to adapt to various downstream tasks with minimal or no supervision. 
From BERT~\cite{devlin2018bert} to GPT-4~\cite{openai2023gpt4}, FMs have fueled ground-breaking advances in natural language tasks. The success of large language models inspired applications to different domains such as speech~\cite{baevski2020wav2vec,radford2022robust}, robotics~\cite{brohan2022rt,stone2023open}, and vision~\cite{kirillov2023segment,zou2023segment}.

Classical methods for medical image segmentation (MIS) implement carefully-customized pipelines (e.g., FreeSurfer~\cite{fischl2012freesurfer}). 
Pipelines might include pre-selecting images that include the region of interest (ROI), preprocessing the images to reduce artifacts and/or noise, and applying image-processing algorithms like thresholding and deformable-templates, with empirically chosen parameters. 
The introduction of deep learning models simplified and improved the performance of automatic segmentation tools~\cite{isensee2021nnu,ronneberger2015u}. 
In deep learning, the common approach involves curating a set of labeled images and training a task-specific model on these data. 
These models can be brittle and not generalize well to new datasets. 
Moreover, they demand the creation of a relatively large labeled training set for each task. 
Importantly, training for each task often requires significant computational resources and expertise.
Recent studies have proposed data augmentation and synthesis methods to address these problems but they are still early stage~\cite{billot2020learning,zhao2019data}.

\begin{figure}[!ht]
    \centering
    \includegraphics[width=\linewidth]{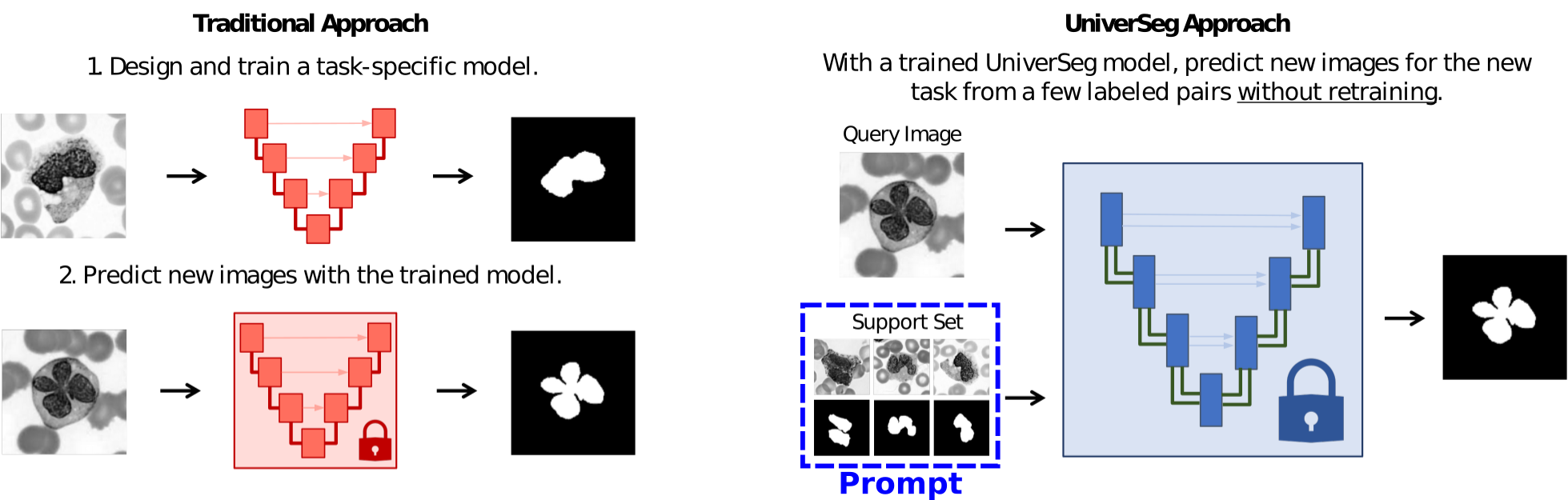}
    \caption{\textbf{Traditional Approach vs. Foundational Model Approach.} Traditional segmentation models like nnUNet are trained first to predict the new images. FMs like UniverSeg and SAM  use a trained model for inference of a new task. Instead of retraining, prompts like support sets are used for UniverSeg and points and masks for SAM (Image modified from \cite{butoi2023universeg}) \label{fig:traditional.vs.foundation}}.
\end{figure}

Recently, several FMs for image segmentation tasks  have been proposed.
These include the Segment Anything Model (SAM) and Segment everything everywhere all at once model (SEEM), which demonstrate great performance in a variety of interactive segmentation tasks in natural images~\cite{kirillov2023segment,zou2023segment}. 
Unlike task-specific models, these FMs are trained with prompt inputs like points and boxes that guide the segmentation tasks. Once trained, these methods solve new tasks without updating their weights (Figure~\ref{fig:traditional.vs.foundation}). 
Another recent FM, UniverSeg~\cite{butoi2023universeg}, is specifically designed to generally solve \textit{medical} image segmentation tasks. 
The ``prompt'' for UniverSeg is a set of image-label pairs, also called a support set.
The support set precisely defines the segmentation task. 
As one of the first FMs developed for medical image segmentation, UniverSeg demonstrated promising performance using limited number of image-label pairs compared to few-shot baseline methods.

A FM for MIS offers several benefits. 
This approach can minimize the need for labeled data, which can represent a significant reduction in cost for developing automatic segmentation tools.
Since these models leverage commonalities across different annotation tasks, adapting a FM to a new task can be made to be computationally efficient and reduce the computational burden for creating task-specific solutions.
Finally, adapting FMs to specific tasks can be made easy and user-friendly, which will  help lower barriers for clinical practitioners to build on these technologies. 

Although promising, studies have shown the limitations of the SAM FM for MIS tasks~\cite{cheng2023sam,deng2023segment,he2023accuracy,hu2023skinsam,huang2023segment,mattjie2023exploring,mazurowski2023segment,roy2023sam,zhou2023can}. 
The inferior performance of SAM on MIS tasks is often attributed to the fact that SAM was trained with natural images.
Some works propose possible remedies, such as prompt-engineering \cite{shi2023generalist,wald2023sam} and fine-tuning \cite{gao2023desam,ma2023segment,wu2023medical} to improve the performance. 
In this paper, we report the potential and limitations of an MIS-specific FM, UniverSeg, by evaluating it for prostate MRI segmentation.

\section{Related Works} 
\subsection{UniverSeg} 
UniverSeg \cite{butoi2023universeg} is a FM for MIS tasks that uses support sets of image-label pairs as a prompt to define new tasks. 
The architecture employs a Cross-Block mechanism leveraging information from the query image and support sets by averaging the feature maps. 
UniverSeg was built using MegaMedical, which contains 53 open-access medical segmentation datasets comprising over 22,000 scans to achieve strong performance when generalizing to held out datasets used to evaluate UniverSeg on unseen anatomies and tasks.

\subsection{Prostate MR Segmentation}
Prostate MR scans have been increasingly acquired as an initial diagnostic tool. The ROI labels are manually segmented for the clinical workflow, for example, biopsy guidance, and surgical/treatment planning. High-quality segmentation labels can be beneficial but the label generation is time-consuming and demands expertise. Thus, automatic segmentation tools can have a large clinical impact.

\section{Experiments}
\subsection{Datasets}
We consider three anatomical ROIs in the prostate that are defined in two datasets. 
For each dataset, we created five sets of support/test splits. 
Since obtaining high-quality ground-truth labels is a significant bottleneck for real-world MIS problems, we focus on the limited sample size scenario. 
We created support sets with randomly selected N=1, 2, 5, and 10 cases, while the other cases were used as test set. 
Since each training case is a 3D volume, we extracted 2D slices from these volumes to create the support or training sets.
Unless specified otherwise, we used 2D slices that contained the ROI. 
All slices are resized to $128\times128$ and intensities are normalized to [0, 1]. 

\subpara{Prostate Gland Segmentation.} We used our in-house prostate MRI dataset (Prostate-Gland) for prostate gland segmentation, amounting to 859 anonymized MRI scans. T2-weighted prostate MRI scans are acquired as part of prostate cancer diagnosis. 

\subpara{Transitional and Peripheral Zone Segmentation.} We used the publicly available zonal anatomy segmentation labels of 204 patients~\cite{cuocolo2021quality}. The transitional zone (TZ) and peripheral zone (PZ) labels are from the training dataset of the PROSTATEx challenge \cite{litjens2014computer} and annotated by expert radiologists, with rigorous quality assessment \cite{cuocolo2021quality}. 
We present two sets of results corresponding to two different labels: PROSTATEx-TZ and PROSTATEx-PZ. 

\subsection{UniverSeg Inference}
One of the crucial limitations of existing FMs for segmentation, including UniverSeg~\cite{butoi2023universeg}, is that they are all trained in 2D. 
However, most medical image segmentation tasks are in 3D, and the ROIs can be present in a small portion of the entire volume.
Thus, many 2D slices will not contain the segmentation label. 
Regular prompt-based FM's like SAM~\cite{kirillov2023segment} struggle with this, as they are expected to return a non-zero result for a given query and prompt.  
Although UniverSeg is trained using 2D slices containing the label, UniverSeg can use images with missing ROIs in the support set, which can be critical for 3D segmentation tasks. 
Following the original paper, in all our experiments, we set the maximum support set size $S$ to 64 2D image-label pairs.
Furthermore, as previously demonstrated, the quality of the result obtained with UniverSeg heavily depends on the quality of the provided support set~\cite{butoi2023universeg}. 
In our experiments, we implement different support set selection strategies, described below. 

\subpara{Slice-index-aware Support Set Selection.} 
The anatomical field-of-view along the z-axis of prostate MR images is roughly similar across subjects.
We leveraged this to implement a support set selection strategy that relies on the slice index $Z$ of the query image. 
For a given query image $I_{q}$, we computed weights for each of the available labeled slices $I_{t}$ as follows: $1 / (|Z_{I_{t}} - Z_{I_{q}}| + 1)$, where $Z_I$ denote the slice index in image $I$. 
Then we randomly selected $S$ annotated slices with a probability proportional to the pre-computed weights.
This is our default support set selection strategy, which was used for the main results.

\subpara{Random Support Set Selection.}
As an ablation, we ignore the z-index and randomly draw $S$ support images from  available labeled slices, where each of these images has the same (uniform) probability.

These support set selection techniques can be restricted to slices where the ROI is present (``ROI-inclusive''), or can consider all possible slices in the training volumes (i.e., be agnostic to whether the ROI is present or absent in the slice, which we refer to as ``ROI-agnostic''). Because UniverSeg was trained with only ``ROI-inclusive'' slices, comparing the result with ``ROI-agnostic'' can serve as a good stress test of the released tool.

\subsection{nnUNet}
As the baseline, we used the (2D) nnUNet, which  trains the model from a random initialization on the given labeled data using heavy data augmentation, automatic network configuration, and ensembling (nnUNet-original)~\cite{isensee2021nnu}. The nnUNet model is widely considered state-of-the-art for a wide range of task-specific segmentation tasks.
For further comparison, we trained and tested the nnUNet model with a smaller network capacity that is similar to the size of the UniverSeg model, which we refer to as nnUNet-small (See Appendix for the details). 

\subsection{Empirical Evaluation}
Because high-performance machines are often unavailable in clinical and medical-research settings, understanding the required computational resources is important to utilize deep learning models for clinical use. As many FMs for segmentation are based on Vision Transformer~\cite{dosovitskiy2020image} trained with large datasets, they involve a large number of parameters. Also, compared to classification problems, MIS models often involve higher memory requirements. 
We performed computational resource analysis on nnUNet and UniverSeg by comparing the number of parameters, training, and inference time. 

As the main performance metric, we used the Dice score~\cite{dice1945measures} that quantifies the overlap between an automatic and ground-truth segmentation, and is widely used in the field.
We compare UniverSeg with nnUNet models, when different number ($N$) of training cases are available. 
We performed ablation studies to understand where the performance improvement occurs for the UniverSeg and nnUNet models. 
We compute Dice both in 2D and in 3D. 
The 2D Dice results are presented only for slices that contain the ROI, and aggregated over all slices in the test subjects.
For these results, we implemented the ROI-inclusive support set strategy.
We also present 3D Dice values, which are computed based on the volumetric overlap in each test subject, which is in turn averaged across subjects.


\section{Results}
\subsection{Computational Resource}
Table~\ref{table:resource} shows computational resources needed for nnUNet and UniverSeg. 
UniverSeg has a much smaller number of parameters and faster inference runtime. Importantly, UniverSeg does not require task-specific training -- saving substantial computational requirement, and obviating the need for a GPU. This substantial savings makes is more applicable to clinical and clinical-research settings.
nnUNet implements five-fold cross-validation, which it in turn uses to ensemble five models. 
This means that for each nnUNet, we store five models and run five inferences.
For nnUNet-orig, the automatic configuration in our experiment yielded models with 20.6M parameters, which is 100 times larger than UniverSeg (1.2M). 
Our nnUNet-small implementation had 1.3M learnable parameters, yet we emphasize that ensembling over cross-validation runs meant that the memory footprint of nnUNet-small is about five times of UniverSeg.
While the inference time for the nnUNet models will not depend on the training set size ($N$), UniverSeg's will, since we need to ensemble over various support sets when $N>2$ for better performance.
However, the support set size does not affect the number of parameters as the Cross-Block of UniverSeg averages the representations of interaction between query and support sets at each step in the network.

\setlength{\tabcolsep}{12pt}
\begin{table}[!ht]\centering
\begin{tabular}{l|ccc}\hline
  & nnUNet--orig. & nnUNet--small & UniverSeg \\
  \hline
\#Params & $20.6 \textrm{ M}\times 5 \textrm{ folds}$ & $1.3 \textrm{ M}\times 5 \textrm{ folds}$ & $\mathbf{1.2}$ \textbf{M} \\
Training time (ms) & $1.6\times 10^{8}$ & $1.2\times 10^{8}$ & -- \\
Inference time (ms) & $9.7\times 10^{3}$ & $7.5\times 10^{3}$ & $\mathbf{6.9\times 10^{2}}$ \\\hline
\end{tabular}
\caption{Computational resource comparison. The values are averaged across ROIs and calculated for N=1 case for all methods.  All models are tested on Nvidia TITAN Xp GPU (12 GB vRAM).\label{table:resource}}
\end{table}

\setlength{\tabcolsep}{6pt}
\begin{table}[ht!]\centering
\scalebox{0.8}{
\begin{tabular}{ll|cccc}\hline
ROI & Method & $N=1$ & $N=2$ & $N=5$ & $N=10$ \\\hline
\multirow{3}{*}{Prostate-Gland} & nnUNet-Orig & $0.592\pm0.088$ & $0.714\pm0.045$ &  $0.810\pm0.007$ & $0.817\pm0.016$ \\

& nnUNet-Small & $0.520\pm0.076$ & $0.698\pm0.057$ &  $0.802\pm0.008$ & $0.808\pm0.019$ \\
\cdashline{2-6}
 
& UniverSeg & $\mathbf{0.711\pm0.008}$ & $\mathbf{0.769\pm0.009}$ &  $0.780\pm0.003$ & $0.802\pm0.005$ \\
\hline
 
\multirow{3}{*}{PROSTATEx-TZ} & nnUNet-Orig & $0.614\pm0.049$ & $0.764\pm0.034$ &  $0.803\pm0.006$ & $0.821\pm0.010$ \\

& nnUNet-Small & $0.599\pm0.066$ & $0.759\pm0.033$ &  $0.800\pm0.006$ & $0.814\pm0.011$ \\
\cdashline{2-6}
 
& UniverSeg & $\mathbf{0.632\pm0.046}$ & $0.717\pm0.010$ &  $0.743\pm0.012$ & $0.754\pm0.015$ \\ \hline

\multirow{3}{*}{PROSTATEx-PZ} & nnUNet-Orig & $0.368\pm0.111$ & $0.589\pm0.041$ &  $0.644\pm0.042$ & $0.706\pm0.018$ \\

& nnUNet-Small & $0.333\pm0.122$ & $0.572\pm0.048$ &  $0.633\pm0.049$ & $0.699\pm0.016$ \\
\cdashline{2-6}
 
& UniverSeg & $\mathbf{0.478\pm0.056}$ & $0.570\pm0.014$ &  $\mathbf{0.647\pm0.018}$ & $0.673\pm0.015$ \\ \hline

\end{tabular}}
\caption{2D Dice scores for UniverSeg and nnUNet models. The scores are averaged across 5 support/test splits. \label{table:performance}}
\end{table}

\subsection{Segmentation Performance}
\begin{figure}[ht!]
    \centering
    \includegraphics[width=\linewidth]{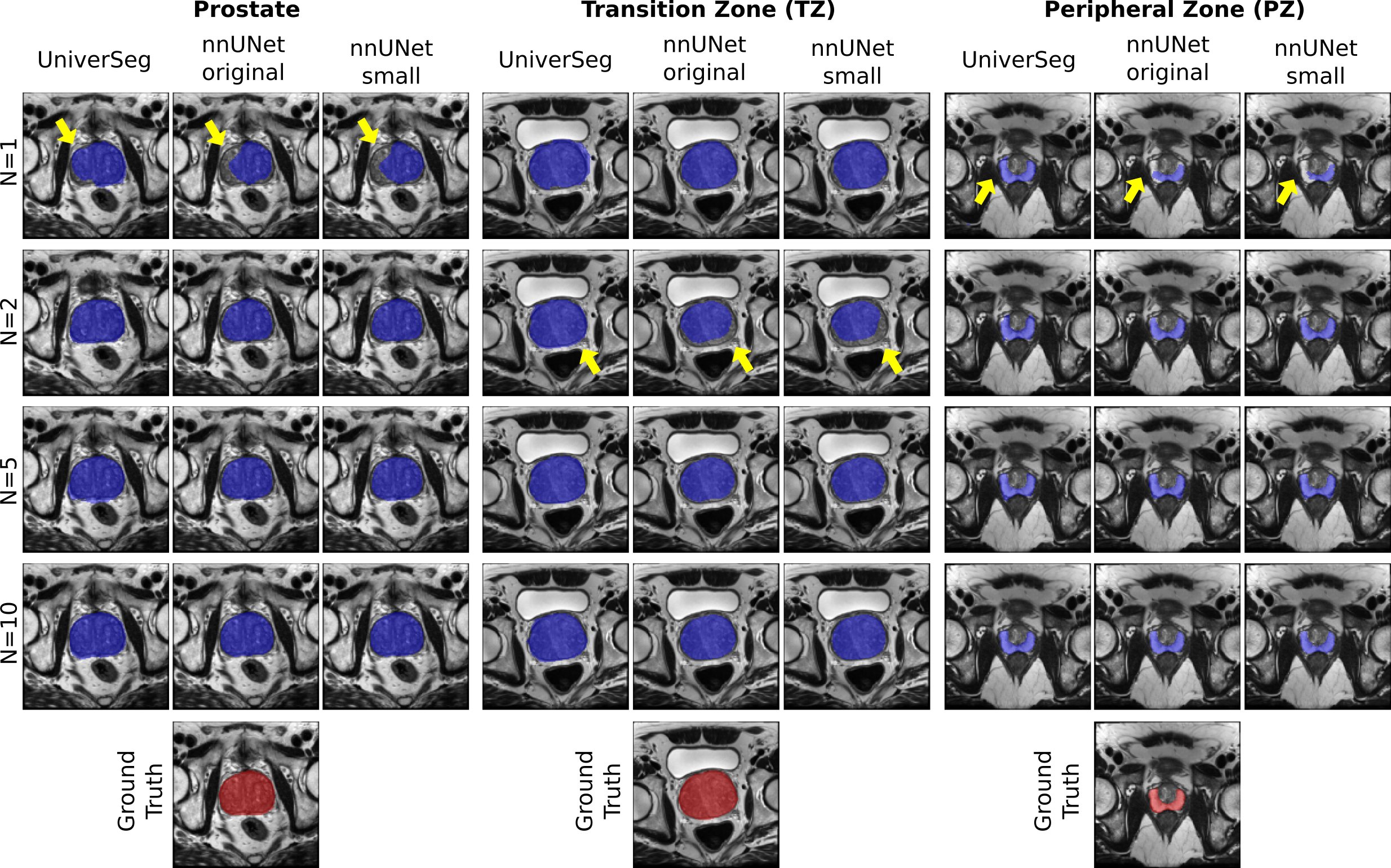}
    \caption{Representative results. UniverSeg results are comparable to the nnUNet baseline. When existing segmentation labels are limited, e.g., $N=1$ and $N=2$, UniverSeg shows superior performance than nnUNet models (highlighted in yellow). \label{fig:performance}}
\end{figure}

We first analyzed segmentation performance for 2D slices that contain the ROI.
Table~\ref{table:performance} and Figure~\ref{fig:performance} show quantitative and qualitative results.
Models perform better when more training images are available. 
For Prostate-Gland segmentation, UniverSeg showed overall comparable results to the nnUNet models, particularly when compared with the size-matched version (nnUNet-small).
Interestingly, UniverSeg achieved good performance given extremely limited annotated data, e.g., $N=1$, outperforming the nnUNet models for all three tasks.
The lower scores in TZ and PZ segmentation have been previously analyzed, and are due to the small size and difficult shape of these ROIs. For example, prior zonal segmentation studies report varying scores ranging between 0.59 to 0.94 showing the difficulty and variability~\cite{bardis2021segmentation,chen2022enhancing,rouviere2022combined,zhu2019fully}. 
The nnUNet models outperform UniverSeg in TZ segmentation when $N=5$ and $N=10$ annotated examples are available. 
This difference is smaller for PZ and only becomes significant at $N=10$. It is important to note that the nnUNet models use test time augmentation, which may improve the UniverSeg performance. 

Table~\ref{table:3d} shows 3D Dice score values and compares two support set selection methods. 
We observe that the ROI-agnostic support selection method which includes slices that are missing the ROI, achieves significantly better results.
This is because, in 3D, there will be many slices that don't include the ROI and if all support examples include the ROI, then the model will likely produce false positive labels for these slices.
This highlights the importance of considering the possibility that the query image might be lacking the ROI.

\setlength{\tabcolsep}{6pt}
\begin{table}[!ht]\centering
\scalebox{0.9}{
\begin{tabular}{ll|cccc}\hline
Support Set Selection &  N & Prostate & PROSTATEx-TZ &  PROSTATEx-PZ \\
\hline
\multirow{4}{*}{ROI-agnostic} 
&  1 & $0.596\pm0.047$ & $0.610\pm0.060$ &  $0.428\pm0.070$ \\
 &  2 & $0.690\pm0.035$ & $0.706\pm0.011$ &  $0.510\pm0.031$ \\
 &  5 & $0.716\pm0.006$ & $0.740\pm0.019$ &  $0.593\pm0.014$ \\
 &  10 & $0.778\pm0.006$ & $0.751\pm0.024$ &  $0.621\pm0.009$ \\
\hline
\multirow{4}{*}{ROI-inclusive} 
&  1 & $0.481\pm0.035$ & $0.579\pm0.066$ &  $0.349\pm0.042$ \\
 &  2 & $0.488\pm0.034$ & $0.665\pm0.009$ &  $0.393\pm0.009$ \\
 &  5 & $0.513\pm0.027$ & $0.685\pm0.016$ &  $0.487\pm0.013$ \\
 &  10 & $0.543\pm0.013$ & $0.707\pm0.019$ &  $0.493\pm0.027$ \\
\hline
\end{tabular}}
\caption{3D Dice scores for UniverSeg models with two different support set selection strategies\label{table:3d}}.
\end{table}

\noindent\textbf{Ablation.} 
We conducted ablation studies for both UniverSeg and nnUNet models to assess the impact of model configuration choices. The nnUNet with the default configurations includes ensembling and test time augmentation. The prediction results from five cross-validation models are ensembled by averaging softmax probabilities and at test time augmentation is applied by mirroring all axis. As the post-processing step did not improve the accuracy on validation sets, we did not post-process the predicted labels. We report the 2D Dice scores of nnUNet models before the ensembling and without the test time augmentation. For UniverSeg, we compared the different slice selection methods. 

Table~\ref{table:ablation} demonstrates the ablation results on prostate gland segmentation. Ensembling gave all models a boost. For nnUNet models, test time augmentation also slightly enhanced the scores. The results of the support set selection methods demonstrate the effect of support set quality. The result of ensembling 5 times with slice-index-aware (z-weighted) selection method showed superior performance than using all images for support sets for both $N=5$ and $N=10$. This, again, highlights the importance of the quality of support sets. The ablation for TZ and PZ achieved the similar results (See Appendix Table 1). 

\setlength{\tabcolsep}{6pt}
\begin{table}[!ht]\centering
\scalebox{0.7}{
\begin{tabular}{ll|cccc}\hline
ROI & Method & $N=1$ & $N=2$ & $N=5$ & $N=10$ \\\hline
\multirow{7}{*}{nnUNet-Orig } & w/o augmentation & $0.590\pm0.085$ & $0.712\pm0.046$ &  $0.809\pm0.007$ & $0.815\pm0.016$ \\
& fold-1 & $0.581\pm0.086$ & $0.681\pm0.060$ &  $0.798\pm0.011$ & $0.808\pm0.017$ \\
& fold-2 & $0.564\pm0.095$ & $0.710\pm0.039$ &  $0.797\pm0.010$ & $0.798\pm0.023$ \\
& fold-3 & $0.590\pm0.092$ & $0.691\pm0.044$ &  $0.795\pm0.014$ & $0.807\pm0.025$ \\
& fold-4 & $0.599\pm0.088$ & $0.708\pm0.043$ &  $0.785\pm0.006$ & $0.804\pm0.006$ \\
& fold-5 & $0.553\pm0.046$ & $0.692\pm0.046$ &  $0.790\pm0.006$ & $0.810\pm0.008$ \\\cdashline{2-6}
& default & $\mathbf{0.592\pm0.088}$ & $\mathbf{0.714\pm0.045}$ &  $\mathbf{0.810\pm0.007}$ & $\mathbf{0.817\pm0.016}$ \\\hline

\multirow{7}{*}{nnUNet-Small } & w/o augmentation & $0.519\pm0.072$ & $0.696\pm0.056$ &  $0.801\pm0.007$ & $0.807\pm0.018$ \\
& fold-1 & $0.537\pm0.047$ & $0.668\pm0.074$ &  $0.784\pm0.014$ & $0.801\pm0.021$ \\
& fold-2 & $0.518\pm0.068$ & $0.686\pm0.051$ &  $0.793\pm0.012$ & $0.792\pm0.023$ \\
& fold-3 & $0.512\pm0.091$ & $0.689\pm0.057$ &  $0.784\pm0.011$ & $0.803\pm0.011$ \\
& fold-4 & $0.508\pm0.076$ & $0.705\pm0.046$ &  $0.787\pm0.015$ & $0.792\pm0.022$ \\
& fold-5 & $0.530\pm0.089$ & $0.680\pm0.045$ &  $0.782\pm0.014$ & $0.798\pm0.020$ \\\cdashline{2-6}
& default & $\mathbf{0.520\pm0.076}$ & $\mathbf{0.698\pm0.057}$ &  $\mathbf{0.802\pm0.008}$ & $\mathbf{0.808\pm0.019}$ \\\hline

\multirow{5}{*}{UniverSeg }  & all & $\mathbf{0.711\pm0.008}$ & $\mathbf{0.769\pm0.009}$ &  $0.778\pm0.006$ & $0.799\pm0.005$ \\ 
& random & -- & -- &  $0.777\pm0.002$ & $0.798\pm0.005$ \\
 & random+5 ensemble & -- & -- &  $0.779\pm0.004$ & $0.800\pm0.006$ \\
 & z-weighted & -- & -- &  $0.777\pm0.002$ & $0.798\pm0.005$ \\\cdashline{2-6}
 & z-weighted +5 ensemble & -- & -- &  $\mathbf{0.780\pm0.003}$ & $\mathbf{0.802\pm0.005}$ \\\hline
\multicolumn{2}{c|}{Average \# of images available for support set} & $14.0\pm2.1$ & $31.4\pm6.5$ & $83.4\pm2.9$ & $148.0\pm3.7$ \\\hline
\end{tabular}}
\caption{2D Dice scores from the ablation study conducted for the prostate segmentation task.\label{table:ablation}}
\end{table}

\subsection{Conclusion}
Based on the successful employment of FMs in multiple domains, we believe FMs will instigate a paradigm shift for medical imaging. In this paper, we evaluated the FM for MIS, called UniverSeg, and discussed its performance and adaptability to prostate segmentation tasks. 

As future directions, we see several limitations and opportunities in a FM for MIS. 
First, FMs for 3D MIS are needed, and promise to be impactful. 
Many medical image data is acquired in 3D and the existing FMs are based on 2D slices extracted from the 3D volumes. 
Previous studies have shown superior performance when designed for 3D compared to 2D data. 
FMs like UniverSeg, where the model can account for images without ROI labels, should be further studied for 3D tasks. 
Second, adaptation of FMs should be further studied. 
Prostate gland and TZ were comparably easier segmentation tasks then the PZ. 
Different approaches would include but not be limited to ensembling  different models, e.g., ensembling nnUNet and UniverSeg results, prompt engineering, and finetuning. 
Third, clinical practitioners can easily adapt FMs in their workflows, as it obviates the need to fine-tune. 
For prostate MRI, some practitioners use an automated prostate gland segmentation tool from the software DynaCAD\footnote{https://www.usa.philips.com/healthcare/product/HC784029/dynacad-prostate}. 
Even though the segmentation needs to be reviewed and edited, the software saves a lot of time over manual segmentation. 
An FM like UniverSeg, can be used for various segmentation tasks even when limited labels are available.

\subsubsection{Acknowledgements} This work was supported by NIH, United States grant R01AG053949 and 1R01AG064027, the NSF, United States NeuroNex grant 1707312, and the NSF, United States CAREER 1748377 grant. 

%
%
%
\bibliographystyle{splncs04}
\bibliography{bibliography}

\begin{thebibliography}{10}
\providecommand{\url}[1]{\texttt{#1}}
\providecommand{\urlprefix}{URL }
\providecommand{\doi}[1]{https://doi.org/#1}

\bibitem{baevski2020wav2vec}
Baevski, A., Zhou, Y., Mohamed, A., Auli, M.: wav2vec 2.0: A framework for
  self-supervised learning of speech representations. Advances in neural
  information processing systems  \textbf{33},  12449--12460 (2020)

\bibitem{bardis2021segmentation}
Bardis, M., Houshyar, R., Chantaduly, C., Tran-Harding, K., Ushinsky, A.,
  et~al.: Segmentation of the prostate transition zone and peripheral zone on
  mr images with deep learning. Radiology: Imaging Cancer  \textbf{3}(3),
  e200024 (2021)

\bibitem{billot2020learning}
Billot, B., Greve, D., Van~Leemput, K., Fischl, B., Iglesias, J.E., Dalca,
  A.V.: A learning strategy for contrast-agnostic mri segmentation. arXiv
  preprint arXiv:2003.01995  (2020)

\bibitem{bommasani2021opportunities}
Bommasani, R., Hudson, D.A., Adeli, E., Altman, R., Arora, S., et~al.: On the
  opportunities and risks of foundation models. arXiv preprint arXiv:2108.07258
   (2021)

\bibitem{brohan2022rt}
Brohan, A., Brown, N., Carbajal, J., Chebotar, Y., Dabis, J., et~al.: Rt-1:
  Robotics transformer for real-world control at scale. arXiv preprint
  arXiv:2212.06817  (2022)

\bibitem{butoi2023universeg}
Butoi, V.I., Ortiz, J.J.G., Ma, T., Sabuncu, M.R., Guttag, J., Dalca, A.V.:
  Universeg: Universal medical image segmentation. arXiv preprint
  arXiv:2304.06131  (2023)

\bibitem{chen2022enhancing}
Chen, C., Qin, C., Ouyang, C., Li, Z., Wang, S., et~al.: Enhancing mr image
  segmentation with realistic adversarial data augmentation. Medical Image
  Analysis  \textbf{82},  102597 (2022)

\bibitem{cheng2023sam}
Cheng, D., Qin, Z., Jiang, Z., Zhang, S., Lao, Q., Li, K.: Sam on medical
  images: A comprehensive study on three prompt modes. arXiv preprint
  arXiv:2305.00035  (2023)

\bibitem{cuocolo2021quality}
Cuocolo, R., Stanzione, A., Castaldo, A., De~Lucia, D.R., Imbriaco, M.: Quality
  control and whole-gland, zonal and lesion annotations for the prostatex
  challenge public dataset. European Journal of Radiology  \textbf{138},
  109647 (2021)

\bibitem{deng2023segment}
Deng, R., Cui, C., Liu, Q., Yao, T., Remedios, L.W., et~al.: Segment anything
  model (sam) for digital pathology: Assess zero-shot segmentation on whole
  slide imaging. arXiv preprint arXiv:2304.04155  (2023)

\bibitem{devlin2018bert}
Devlin, J., Chang, M.W., Lee, K., Toutanova, K.: Bert: Pre-training of deep
  bidirectional transformers for language understanding. arXiv preprint
  arXiv:1810.04805  (2018)

\bibitem{dice1945measures}
Dice, L.R.: Measures of the amount of ecologic association between species.
  Ecology  \textbf{26}(3),  297--302 (1945)

\bibitem{dosovitskiy2020image}
Dosovitskiy, A., Beyer, L., Kolesnikov, A., Weissenborn, D., Zhai, X., et~al.:
  An image is worth 16x16 words: Transformers for image recognition at scale.
  arXiv preprint arXiv:2010.11929  (2020)

\bibitem{fischl2012freesurfer}
Fischl, B.: Freesurfer. Neuroimage  \textbf{62}(2),  774--781 (2012)

\bibitem{gao2023desam}
Gao, Y., Xia, W., Hu, D., Gao, X.: Desam: Decoupling segment anything model for
  generalizable medical image segmentation. arXiv preprint arXiv:2306.00499
  (2023)

\bibitem{he2023accuracy}
He, S., Bao, R., Li, J., Grant, P.E., Ou, Y.: Accuracy of segment-anything
  model (sam) in medical image segmentation tasks. arXiv preprint
  arXiv:2304.09324  (2023)

\bibitem{hu2023skinsam}
Hu, M., Li, Y., Yang, X.: Skinsam: Empowering skin cancer segmentation with
  segment anything model. arXiv preprint arXiv:2304.13973  (2023)

\bibitem{huang2023segment}
Huang, Y., Yang, X., Liu, L., Zhou, H., Chang, A., et~al.: Segment anything
  model for medical images? arXiv preprint arXiv:2304.14660  (2023)

\bibitem{isensee2021nnu}
Isensee, F., Jaeger, P.F., Kohl, S.A., Petersen, J., Maier-Hein, K.H.: nnu-net:
  a self-configuring method for deep learning-based biomedical image
  segmentation. Nature methods  \textbf{18}(2),  203--211 (2021)

\bibitem{kirillov2023segment}
Kirillov, A., Mintun, E., Ravi, N., Mao, H., Rolland, C., et~al.: Segment
  anything. arXiv preprint arXiv:2304.02643  (2023)

\bibitem{litjens2014computer}
Litjens, G., Debats, O., Barentsz, J., Karssemeijer, N., Huisman, H.:
  Computer-aided detection of prostate cancer in mri. IEEE transactions on
  medical imaging  \textbf{33}(5),  1083--1092 (2014)

\bibitem{ma2023segment}
Ma, J., Wang, B.: Segment anything in medical images. arXiv preprint
  arXiv:2304.12306  (2023)

\bibitem{mattjie2023exploring}
Mattjie, C., de~Moura, L.V., Ravazio, R.C., Kupssinsk{\"u}, L.S., Parraga, O.,
  et~al.: Exploring the zero-shot capabilities of the segment anything model
  (sam) in 2d medical imaging: A comprehensive evaluation and practical
  guideline. arXiv preprint arXiv:2305.00109  (2023)

\bibitem{mazurowski2023segment}
Mazurowski, M.A., Dong, H., Gu, H., Yang, J., Konz, N., Zhang, Y.: Segment
  anything model for medical image analysis: an experimental study. arXiv
  preprint arXiv:2304.10517  (2023)

\bibitem{openai2023gpt4}
OpenAI: Gpt-4 technical report (2023)

\bibitem{radford2022robust}
Radford, A., Kim, J.W., Xu, T., Brockman, G., McLeavey, C., Sutskever, I.:
  Robust speech recognition via large-scale weak supervision. arXiv preprint
  arXiv:2212.04356  (2022)

\bibitem{ronneberger2015u}
Ronneberger, O., Fischer, P., Brox, T.: U-net: Convolutional networks for
  biomedical image segmentation. In: MICCAI 2015: 18th International
  Conference, Munich, Germany, October 5-9, 2015, Proceedings, Part III 18. pp.
  234--241. Springer (2015)

\bibitem{rouviere2022combined}
Rouvi{\`e}re, O., Moldovan, P.C., Vlachomitrou, A., Gouttard, S., Riche, B.,
  et~al.: Combined model-based and deep learning-based automated 3d zonal
  segmentation of the prostate on t2-weighted mr images: clinical evaluation.
  European Radiology pp. 1--12 (2022)

\bibitem{roy2023sam}
Roy, S., Wald, T., Koehler, G., Rokuss, M.R., Disch, N., et~al.: Sam. md:
  Zero-shot medical image segmentation capabilities of the segment anything
  model. arXiv preprint arXiv:2304.05396  (2023)

\bibitem{shi2023generalist}
Shi, P., Qiu, J., Abaxi, S.M.D., Wei, H., Lo, F.P.W., Yuan, W.: Generalist
  vision foundation models for medical imaging: A case study of segment
  anything model on zero-shot medical segmentation. Diagnostics
  \textbf{13}(11), ~1947 (2023)

\bibitem{stone2023open}
Stone, A., Xiao, T., Lu, Y., Gopalakrishnan, K., Lee, K.H., et~al.: Open-world
  object manipulation using pre-trained vision-language models. arXiv preprint
  arXiv:2303.00905  (2023)

\bibitem{wald2023sam}
Wald, T., Roy, S., Koehler, G., Disch, N., Rokuss, M.R., et~al.: Sam. md:
  Zero-shot medical image segmentation capabilities of the segment anything
  model. In: Medical Imaging with Deep Learning, short paper track (2023)

\bibitem{wu2023medical}
Wu, J., Fu, R., Fang, H., Liu, Y., Wang, Z., et~al.: Medical sam adapter:
  Adapting segment anything model for medical image segmentation. arXiv
  preprint arXiv:2304.12620  (2023)

\bibitem{zhao2019data}
Zhao, A., Balakrishnan, G., Durand, F., Guttag, J.V., Dalca, A.V.: Data
  augmentation using learned transformations for one-shot medical image
  segmentation. In: Proceedings of the IEEE/CVF conference on computer vision
  and pattern recognition. pp. 8543--8553 (2019)

\bibitem{zhou2023can}
Zhou, T., Zhang, Y., Zhou, Y., Wu, Y., Gong, C.: Can sam segment polyps? arXiv
  preprint arXiv:2304.07583  (2023)

\bibitem{zhu2019fully}
Zhu, Y., Wei, R., Gao, G., Ding, L., Zhang, X., et~al.: Fully automatic
  segmentation on prostate mr images based on cascaded fully convolution
  network. Journal of Magnetic Resonance Imaging  \textbf{49}(4),  1149--1156
  (2019)

\bibitem{zou2023segment}
Zou, X., Yang, J., Zhang, H., Li, F., Li, L., et~al.: Segment everything
  everywhere all at once. arXiv preprint arXiv:2304.06718  (2023)

\end{thebibliography}
\newpage
\includepdf[pages=-]{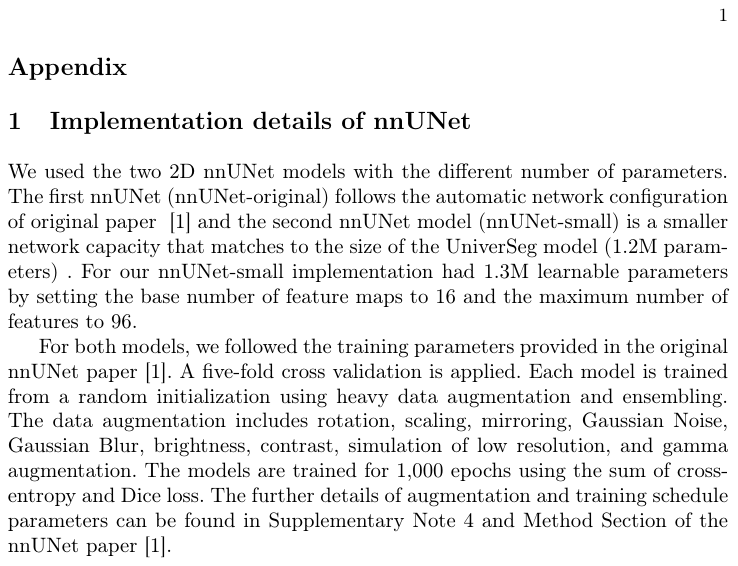}
\end{document}